# Prosocial and Financial Incentives for Biodiversity Conservation:

# A Field Experiment Using a Smartphone App

**Shusaku Sasaki[a,*], Takahiro Kubo[b], Shodai Kitano[c]**

[a] Center for Infectious Disease Education and Research (CiDER), Osaka University

[b] National Institute for Environmental Studies; Interdisciplinary Centre for Conservation Science, University of Oxford

[c] Graduate School of Economics, Osaka University

[*]**Corresponding author:**

Shusaku Sasaki

ssasaki.econ@cider.osaka-u.ac.jp

**Author contributions**

SS conceived and coordinated the study with TK. SS contributed to supervision and funding acquisition. SS and TK developed the design of the field experiment and implemented it. SS analyzed the data with SK. SS wrote the manuscript with contributions from TK and SK. SS, TK, and SK contributed to reviewing, and editing.

**Funding**

Sasaki was supported for this study by Center for Infectious Disease Education and Research

(CiDER), Osaka University, and the Japan Society for the Promotion of Science [grant number: 20H05632].

**Declaration of competing interest**

The authors declare that they have no known competing financial interests or personal relationships that could have appeared to influence the work reported in this paper.

**Ethical considerations**

We obtained ex-ante approval from the ethics committee of Osaka University, Japan (2022CRER0901). We also registered the experimental design with the AEA RCT Registry (Sasaki and Kubo, 2022).

# Abstract

Ascertaining the number, type, and location of plant, insect, and animal species is essential for biodiversity conservation: however, comprehensively monitoring the situation only through public fixed-point surveys is challenging, and therefore information voluntarily provided by citizens assists in ascertaining the species distribution. To effectively encourage the citizens' data sharing behavior, this study proposed a prosocial incentive scheme in which, if they provide species information, donations are made to activities for saving endangered species. We conducted a field experiment with users (N=830) of a widely-used Japanese smartphone app to which they post species photos and measured the incentive's effect on their posting behavior. In addition, we measured the effect of a financial incentive scheme that provides monetary rewards for posting species photos and compared the two incentives' effects. The analyses revealed that while the prosocial incentive did not increase the number of posts on average, it did change the contents of posts, increasing the proportion of posts on rare species. On the contrary, the financial incentive statistically significantly increased the number of posts, in particular, on less rare and invasive species. Our findings suggest that the prosocial and financial incentives could stimulate different motivations and encourage different posting behaviors.

# 1. Introduction

Biodiversity loss is proceeding at a pace never seen before in the human history (IPBES, 2024). For effective biodiversity conservation, gathering and maintaining precise scientific data to ascertain locations of biodiversity loss, habitats of rare and invasive species, and changes in land use is essential. Since it is difficult for expert surveys to fully carry out the data collection and maintenance, citizen science has recently played a significant role. Citizens worldwide voluntarily provide information to help understand biodiversity loss, distribution of species, and land use changes (Atsumi et al., 2023; Eichenberg et al., 2020; Liu et al., 2022; Negrete et al., 2020).

However, citizen science data faces challenges. Generally, citizens report information based on personal preferences, potentially making the reported data biased. For instance, studies have shown that citizen reports increase on warm days and during weekends and holidays compared to cold days and weekdays (Bas et al., 2008; Brum-Bastos et al., 2018; Courter et al., 2013; Sparks et al., 2008). This makes policymakers cautious about using citizen science data directly.

Research has explored various methods to address the challenges with citizen science data. Some studies have focused on statistically adjusting these biases (Bird et al., 2014; Cameron and Kolstoe, 2022; Van Strien et al., 2013), whereas others have investigated the benefits of merging citizen science and expert survey data (Bradter et al., 2018). Furthermore, recent works, including Diekert et al. (2023), have started to explore interventions aimed at guiding citizen behavior towards more desirable information reporting of plant, insect, and animal species.

Biodiversity is a type of public good (Perrings and Gadgil, 2003). Thus, contributing species distribution data for conservation is also a type of public good provision. Economically, it is well-known that public goods with positive externalities often face under-supply, falling short of socially optimal levels (Cornes and Sandler, 1996). To address this problem, financial incentives have been widely used to increase supply in various fields (Barber and West, 2022).[1] However, the effectiveness of financial incentives in promoting prosocial behavior, which is often driven by intrinsic motives, remains unclear. Although traditional economics predict that financial incentives could promote prosocial behavior by changing costs, behavioral economics suggests that financial incentives might diminish intrinsic motivation, reducing people's contributions (Bénabou and Tirole, 2003; Gneezy et al., 2011; Handberg and Angelsen, 2019).

Diekert et al. (2023) used non-experimental data and a difference-in-differences (DID) method and revealed that financial incentives, such as contest prizes, increased information posting on species to an online platform for biodiversity conservation among people in Germany. However, they also found a reduction in diversity of reported species information, which may indicate a tendency to focus on a specific species to obtain monetary rewards.

Our study uses a controlled field experiment and adds new insights to this emerging research stream on intervention strategies. We introduce two interventions to enhance species information reporting, focusing on the conditions where it typically decreases. The first intervention, inspired by Diekert et al. (2023), is providing financial incentives, while the second involves making donations to biodiversity conservation activities in response to experimental participants' posts. We evaluate the effectiveness of these interventions through a randomized controlled trial on a Japan's smartphone app widely-used for posting species information on plants, insects, and animals.

---

[1] Other interventions are non-financial ones called nudges. Nudges encourage people's desirable behavior by changing the environmental setting in which they make choices and framing of the messages they receive. In the context of biodiversity conservation, nudges have been applied to fundraising activities (Kubo et al., 2018; Kubo et al., 2023). However, in general, the effects of nudges were reported to be small (DellaVigna and Linos, 2022).

One unique feature of our study is newly adding donation-related incentives, which we call prosocial incentives. These have been studied in laboratories (Imas, 2014) and applied to charitable giving in the real world as "matching gifts" (Epperson and Reif, 2019; Sasaki et al., 2022). Under the matching gift scheme, people can deliver more than the amount they choose to donate, which is expected to increase their willingness to donate and raise the amount of out-of-pocket donations. Recently, prosocial incentives have been applied to contexts other than charitable giving, such as encouraging physical activities (Galárraga et al., 2020; Harkins et al., 2017; Sumida et al., 2014; Yuan et al., 2021). Our prosocial incentive treatment is a new application in the realm of biodiversity conservation. Under this treatment, people can simultaneously gain intrinsic utility from plant, insect, or animal species postings and intrinsic utility from charitable donations. Both providing data on species information and donating to endangered species conservation efforts can contribute to biodiversity conservation, and these actions are complementary rather than substitutes. As people's utility gains are enhanced by simultaneously engaging in complementary behaviors (Adena and Huck, 2017), this method is predicted to facilitate people's species posting behaviors.

The analyses show the results suggesting that the prosocial and financial incentives could stimulate different motivations and promote different posting behaviors. The prosocial incentive does not increase the number of species posts on average. In contrast, the financial incentive has a significant impact on increasing the number of posts. However, we find another tendency when looking at contents of the posted species information. The prosocial incentive increases the proportion of posts of rare species, while the financial incentive increases in particular the number of posts of less rare and invasive species.

The structure of this paper is outlined as follows: **Section 2** details the design of the field experiment. **Section 3** presents the basic analysis results, followed by **Section 4** which delves into the further analysis. The paper discusses and concludes in **Section 5**.

## 2. Field Experiment

### 2.1. Experimental Design

We conducted the field experiment from September to November 2022 in Japan, in partnership with a private company offering a smartphone application (app) called "biome." On this app, Japanese users can post photos of wildlife they encounter, including plants, insects, and animals, and AI identifies the species information (Atsumi et al., 2023). The app records species information, location, and time data for each photo. Thus, it enables the users to create a catalog of the species they encounter. The app gives users non-financial points for posting species photos, and they would enjoy accumulating them in order to reach a higher level, similar to computer games. Launched in April 2019, biome had 630,000 users as of September 2022. The company frequently partners with the Ministry of the Environment, local governments, and other organizations to conduct surveys to ascertain species' distributions. Similarly, we conducted this project in discussion with Japan Conference for 2030 Global Biodiversity Framework, the Ministry of the Environment, Government of Japan.

We recruited Japanese participants from September 15 to 30, 2022, through an in-app advertisement. We obtained their opt-in consent for participation and usage of data recorded in the app and then asked them to answer a questionnaire survey on information such as their socio-economic attributes, app usage insights, etc.[2] We explained to the participants that this project, titled Posting Marathon Campaign,[3] aimed to record people's species posting behaviors and understanding their preferences, contributing to biodiversity conservation. Participants were asked to take and post photos of plant, insect, or animal species in their spare time during this project. Participants who completed the project received a 300 JPY gift certificate for online shopping as a basic reward.[4]

From the initial 887 applications, we excluded those with ID mismatches, duplicate applications, and other issues. This left us with 830 participants for the analysis. We divided the participants into three groups: control group, prosocial incentive treatment group, and financial incentive treatment group. We performed stratified randomization for the assignment based on the number of species photo posts during the two-week baseline period before the experiment (the first half of September) and the participants' motivations ascertained through the survey.

---

[2] Although the external validity of field experiments with informed consent is generally lower than that of natural field experiments without it, the former has the advantages of fewer ethical concerns and possibility of merging naturally occurring data with self-reported survey data for analysis (Harrison and List, 2004). Such field experiments have often been employed in ecological economics (Kerr et al., 2012)

[3] As the term "experiment" has been reported to be offensive and confusing to the general public (Samek, 2019), we named this project the "Posting Marathon Campaign."

[4] We determined the amount of the reward after a discussion with the app company, considering the non-commercial nature of this app. As of November 2022, one USD equaled 142.62 JPY.

Our two incentive schemes are aimed to increase species posts by users, particularly during periods when posts typically decline, such as during the fall rather than summer and weekdays rather than weekends and holidays. We set a two-week treatment period (October 17–21 and 24–28) and communicated the following information to the participants:

**Control group (n=275):** Participants assigned to this group were asked to take and post photos of species in their spare time on weekdays.

**Prosocial incentive treatment group (n=273):** Participants assigned to this group were asked to take and post photos of species in their spare time on weekdays. In addition, based on the number of species photos posted by each participant during the five weekdays, a donation was made to an activity related to biodiversity conservation.

*"For each different species photo you post during these five weekdays, the campaign office will donate 25 JPY to the endangered species protection activities. If you post photos of 10 different species, the office will donate a maximum of 250 JPY. The donation will be delivered to the Japan Wildlife Conservation Society's Activities to Protect Endangered Species in Japan."*

**Financial incentive treatment group (n=282):** Participants assigned to this group were asked to take and post photos of species in their spare time on weekdays. In addition, based on the number of species photos posted by each participant during the five weekdays, they received gift certificates that could be used for online shopping.

> *"For each different species photo you post during these five weekdays, you will receive an Amazon gift certificates (e-mail type) of 25 JPY. If you post photos of 10 different species, you will receive a maximum of 250 JPY gift certificates. Amazon gift certificates can be used for online shopping at Amazon."*

A key aspect of the two treatments is that we provided incentives based on the number of posted photos of unique, different species (i.e., posted photos of the same species were not counted) rather than the number of total posted species photos. Participants in the treatment groups received prosocial or financial incentives in each of the first and second treatment weeks, and thus the maximum amount of the incentive they could receive was 500 JPY. We communicated the above message via e-mail on October 11 before the treatment period, and on October 17 and 24 during the treatment period. **Appendix A** shows the experimental schedule. **Supplementary Materials** contains English translations of the original Japanese messages.

Before starting the experiment, we obtained approval for our research project from the Ethics Committee at Osaka University (2022CRER0901). We also pre-registered the experimental design with the AEA RCT Registry (Sasaki & Kubo, 2022).

## 2.2. Analytical Procedure

We received the data on posted species photos and information for each user from the app company. Using the data, we constructed two outcomes:

i. Number of posted photos of unique, different species on weekdays;

ii. Number of posted photos of species on weekdays (including multiple photos of the same species).

The first outcome counts the number of postings of unique, different species photos. Thus, this is different from the second outcome, which counts even if a participant posts multiple photos of the same species.

We examined the treatment effects by applying a Difference-in-Differences (DID) fixed estimation approach to the field experimental panel data. The baseline period was September 5–9 and 12–16, and the treatment period was October 17–21 and 24–28.[5] We assumed Strongly Ignorable Treatment Assignment and Stable Unit Treatment Values Assumption (SUTVA). We estimated the following equation:

$$Posts_{it} = \alpha_0 + \alpha_1 Prosocial_i \times Week1_t + \alpha_2 Prosocial_i \times Week2_t \\ + \alpha_3 Financial_i \times Week1_t + \alpha_4 Financial_i \times Week2_t + \alpha_5 Week1_t \\ + \alpha_6 Week2_t + \varepsilon_{it}, \quad (1)$$

where $i$ and $t$ are participants and week, respectively; $Posts_{it}$ is the participant $i$'s number of photo posts of unique, different species during five weekdays in week $t$.

---

[5] As we used the means for September 5–9 and September 12–16 for the baseline period data, we obtained three time points per participant: baseline, first treatment period, and second treatment period. Thus, the total numbers of participants and observations in our analysis sample were 830 and 2,490, respectively.

$Prosocial_i$ is 1 for the prosocial incentive treatment group, and 0 otherwise. $Financial_i$ is 1 for the financial incentive treatment group, and 0 otherwise. $Week1_t$ ($Week2_t$) is 1 in the first (second) week of the treatment period, October 17–21 (24–28), and 0 otherwise. $\varepsilon_{it}$ is an error term.

We used standard errors clustered at each participant and account for possible within-individual serial correlation. For a robustness check, we additionally used weather information for prefectures (weekly average temperature, rainfall, and hours of sunshine) and controlled for temporal changes in situations and environments in which participants take and post photos.

The coefficients of interest were $\alpha_1$, $\alpha_2$, $\alpha_3$, and $\alpha_4$, respectively. $\alpha_1$ and $\alpha_2$ captured the difference between the prosocial incentive treatment and control groups in terms of changes in $Posts$ from the baseline to the treatment period. $\alpha_3$ and $\alpha_4$ captured this difference between the financial incentive treatment and control groups.

## 2.3. Balance Check

We confirmed the homogeneity between the three groups. First, as shown in **Table 1**, no statistically significant differences were found between the groups for the average numbers of two outcomes at the baseline period. Second, we found no statistically significant differences for age, gender, having a cohabiting partner, educational years, and economic level.[6] We found no statistically significant differences in free time (non-working hours)[7] or reason for participating in this campaign (degree of prosocial motivation).[8]

**Table 1.** Descriptive statistics

|  | Control Group n=275 | | Prosocial Incentive Treatment Group n=273 | | Financial Incentive Treatment Group n=282 | |
|---|---|---|---|---|---|---|
|  | Mean | S.D. | Mean | S.D. | Mean | S.D. |
| Number of posts of unique, different species photos on weekdays (baseline) | 5.602 | 12.53 | 4.907 | 10.259 | 5.399 | 11.701 |
| Number of posts of species photos on weekdays (baseline) | 6.502 | 15.804 | 5.342 | 11.318 | 6.282 | 14.963 |
| Age | 45.825 | 11.685 | 47.421 | 12.135 | 47.004 | 12.358 |
| Female (1/0) | 0.535 | 0.5 | 0.498 | 0.501 | 0.521 | 0.5 |
| Having a cohabiting partner (1/0) | 0.687 | 0.464 | 0.733 | 0.443 | 0.759 | 0.429 |
| Educationcal years | 15.269 | 2.259 | 15.234 | 2.129 | 15.255 | 2.265 |
| Economic level (Ability to pay) | 7.71 | 0.56 | 7.71 | 0.659 | 7.653 | 0.588 |
| Free hours excluding working hours per week | 135.022 | 19.788 | 134.194 | 20.016 | 134.138 | 20.354 |
| Degree of prosocial motivation for participation to the campaign (from 2 to 10) | 6.927 | 1.407 | 6.949 | 1.442 | 6.901 | 1.378 |

---

[6] After a discussion with the app company, we decided not to ask the participants' income directly. We asked, "What is the maximum amount you would be willing to pay for lunch?" The answers were used as a variable for their economic level.

[7] We asked the participants, "How many days per week do you work?" and "How many hours per day, on average, do you work?" We calculated the length of working hours per week based on their answers. We then calculated the length of free hours from which the working hours were excluded.

[8] We used the following statements to determine the participants' reasons for participating in the campaign: "I think that participating in this campaign will contribute to biodiversity conservation and academic research" and "I can receive gift certificates by participating in this campaign." The participants were asked to rate to what extent they agreed with the statements on a five-point Likert scale (1 = strongly disagree; 5 = strongly agree). We inverted the latter item and added the scores of the two items to obtain an indicator of the prosocial degree (from 2 to 10).

## 3. Basic Analysis: Effects on Number of Posts

### 3.1. Descriptive and Estimation Results

We estimated the treatment effects of prosocial and financial incentives on posts of species photos. Our findings revealed that the prosocial incentive did not enhance either total or unique, different species posting frequency, whereas the financial incentive significantly increased both. The positive effect of the financial incentive was pronounced among participants who had a moderate frequency of unique, different species posts before the experiment. In addition, we found that the increase due to the financial incentive occurred only during weekdays within the treatment period, with no effect during the weekends or after the treatment period ended.

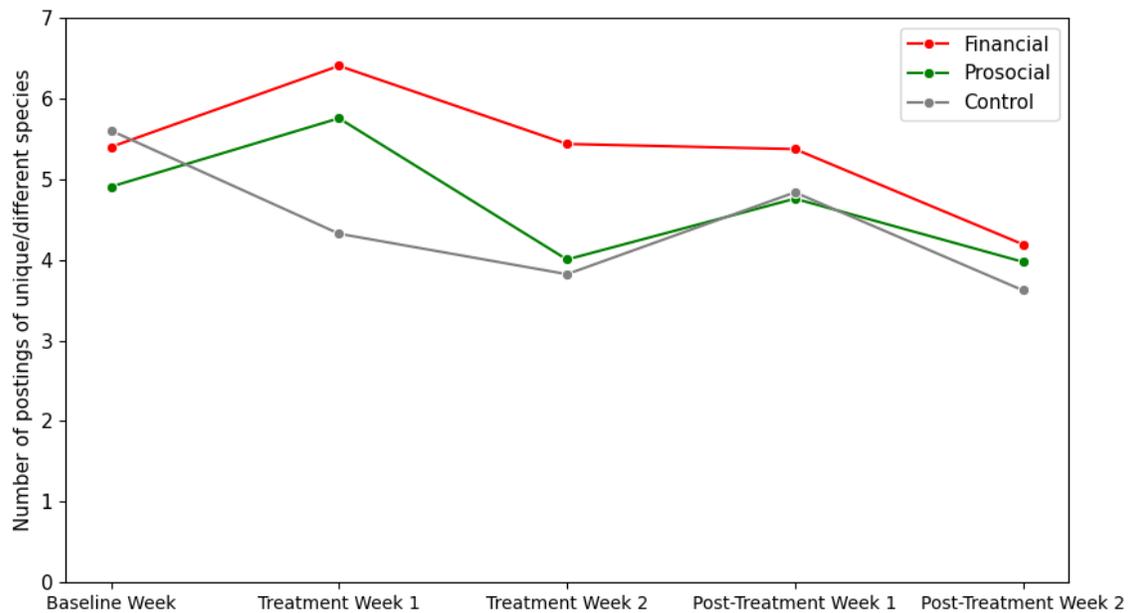

**Figure 1.** Number of Posts of Unique, Different Species

Figure 1 shows the time trends in the number of posted photos of unique, different species on weekdays (i.e., the primary outcome) for each group. The results demonstrated that the number of posts had similar values for each group during the baseline period. The number in the control group dropped during the treatment period but increased in the financial incentive group. In the prosocial incentive group, the number of posts increased during the first treatment week but decreased during the second week. After the treatment period ended, all groups had similar values.

**Table 2.** Basic Analysis: Effects on Number of Posts

|  | (1) | (2) | (3) | (4) |
|---|---|---|---|---|
| Analysis sample: | Total | | | |
| Number of participants: | n=830 | | | |
| Number of observations: | n=2,490 | | | |
| Dependent variable: | Number of posts of unique, different species photos | | Number of posts of species photos | |
| Prosocial incentive*Treatment week 1 | 2.126 | 2.116 | 2.948 | 2.933 |
|  | (1.599) | (1.592) | (2.129) | (2.119) |
| Prosocial incentive*Treatment week 2 | 0.877 | 0.889 | 0.240 | 0.256 |
|  | (0.844) | (0.840) | (1.165) | (1.172) |
| Financial incentive*Treatment week 1 | 2.284** | 2.229** | 2.476** | 2.397** |
|  | (0.984) | (0.984) | (1.127) | (1.128) |
| Financial incentive*Treatment week 2 | 1.817** | 1.753** | 0.975 | 0.873 |
|  | (0.842) | (0.844) | (1.161) | (1.176) |
| Treatment week 1 | -1.278* | -7.007 | -1.396* | -9.582 |
|  | (0.680) | (6.252) | (0.750) | (8.541) |
| Treatment week 2 | -1.780*** | -9.283 | -1.051 | -11.737 |
|  | (0.643) | (8.127) | (0.989) | (11.154) |
| Time-varing attributes | NO | YES | NO | YES |
| Constant term | 5.304*** | 22.671 | 6.046*** | 30.638 |
|  | (0.260) | (17.745) | (0.316) | (24.244) |

Notes: We use standard errors clustered at each participant; * $p < 0.1$, ** $p < 0.05$, *** $p < 0.01$

**Table 2, Column 1** presents the regression results for posted photos of unique, different species. The prosocial incentive group showed a coefficient of 2.126 in the first treatment week, which was not statistically significant. The effect size decreased in the second treatment week. The financial incentive group yielded coefficients of 2.284 in week 1 and 1.817 in week 2, both statistically significant ($p<.05$). However, a direct comparison of the prosocial and financial incentive groups revealed no significant differences, likely due

to the large standard errors of the prosocial incentive.

**Column 3** presents the results for total number of species photo posts. The financial incentive group exhibited an increase in posts only during the first treatment week (2.475, $p<.05$).

For robustness check, **Columns 2 and 4** additionally controlled for weather data (weekly average temperature, rainfall, and sunshine hours) for the participants' locations. This considers the changing conditions and environments in which they take and post photos. The results showed that the size and statistical significance remained largely unchanged.

**Table 3.** Basic Analysis: Heterogeneity

| | (1) | (2) | (3) | (4) | (5) | (6) |
|---|---|---|---|---|---|---|
| Analysis sample: | SMALL | | MEDIUM | | LARGE (over 11) | |
| | (Number of posts of unique, different species photos at the baseline period) | | | | | |
| Number of participants: | n=375 | | n=344 | | n=111 | |
| Number of observations: | n=1,125 | | n=1,032 | | n=333 | |
| Dependent variable: | Number of posts of unique, different species photos | | | | | |
| Prosocial incentive*Treatment week 1 | 0.013 | -0.017 | 3.449 | 2.941 | 4.716 | 4.705 |
| | (0.568) | (0.563) | (3.240) | (2.768) | (5.023) | (4.864) |
| Prosocial incentive*Treatment week 2 | -0.138 | -0.163 | 0.321 | -0.208 | 5.915 | 5.779 |
| | (0.601) | (0.595) | (0.670) | (0.828) | (4.855) | (4.689) |
| Financial incentive*Treatment week 1 | 1.018 | 1.026 | 3.447** | 2.977** | 4.057 | 3.826 |
| | (0.639) | (0.635) | (1.422) | (1.428) | (4.768) | (4.761) |
| Financial incentive*Treatment week 2 | 0.445 | 0.472 | 2.000*** | 1.516* | 6.760 | 6.389 |
| | (0.665) | (0.661) | (0.729) | (0.857) | (4.457) | (4.401) |
| Treatment week 1 | 1.561*** | 2.658 | -0.188 | -14.440 | -14.444*** | -32.835** |
| | (0.448) | (1.679) | (0.479) | (13.157) | (4.119) | (15.348) |
| Treatment week 2 | 1.332*** | 2.733 | -0.504 | -18.994 | -16.472*** | -41.286** |
| | (0.513) | (2.097) | (0.455) | (17.155) | (3.482) | (20.605) |
| Time-varing attributes | NO | YES | NO | YES | NO | YES |
| Constant term | 0.119 | -3.794 | 3.654*** | 42.270 | 27.937*** | 95.594** |
| | (0.146) | (4.418) | (0.424) | (37.808) | (1.108) | (42.564) |

Notes: We use standard errors clustered at each participant; * $p < 0.1$, ** $p < 0.05$, *** $p < 0.01$

In **Table 3**, we examined how each treatment effect varies with the pre-experiment frequency of unique, different species postings. We categorized the participants into three groups, including Small (the mean is 0.119), Medium (the mean is 3.654), and Large (the mean is 27.937).[9] The findings indicated that the effect of financial incentives on increasing unique, different species posts was primarily observed in the Medium group. The treatments' coefficients, including the one for the prosocial incentive, for the Large group were large but not statistically significant due to their large standard errors.[10]

As shown in **Tables Appendix B.1** and **B.2**, the increase in unique, different species posts due to the financial incentive was limited to weekdays during the treatment period. This effect did not continue on the weekends or after the treatment period.

**3.2. Interpretation**

The result of **Table 2, Column 1** suggests a substantial impact of the financial incentive on the number of photo posts of unique, different species. Initially, its number averaged 5.304, according to the constant term. During the first treatment week, there was a decrease in the control group from 5.304 to 4.026 (a drop of 1.278). However, the financial incentive reversed this trend, increasing the number of posts by around 2.284 (from 4.026 to 6.310), or +56.7%. A similar pattern occurred in the second treatment week, with a rise of 51.6%, highlighting the effectiveness of financial incentives in boosting posts of unique, different species photos.

Considering the effects of the financial incentives on total and unique, different species posting frequency in **Table 1**, we interpret the results as follows: In the first treatment week, the financial incentive led to an increase in total and unique, different species posts. However, in the second week, the total number of species posts did not increase, whereas that of unique, different species posts increased. This implies that participants became selective in their posts, likely influenced by their experiences in the first week.

---

[9] The maximum number of unique, different species for which donation or monetary incentives were granted was set at 10. Therefore, we categorized the participants with over 11 pre-experiment postings as the Large group and then divided the rest of the participants into two equal groups (Medium and Small).
[10] Although we performed a subsample analysis using the attribute variables included in the balance check, we did not find any significant heterogeneity.

The results of **Tables B1** and **B2** in **Appendix B** imply that the increase in the number of unique, different species posts disappeared on the weekends and after the treatment period. This tendency has been observed in previous studies on financial and prosocial incentives in other contexts (Barte and Wendel-Vos, 2017; Epperson and Reif, 2019). These results imply that no "boomerang effect" existed; species posts did not decrease on weekends or after the treatment period.[11] As this effect is a concern in behavioral intervention studies (Allcott, 2011; Scharf et al., 2022), the absence of this effect in our study is meaningful.

## 4. Further Analysis: Effects on Contents of Posts

### 4.1. Estimation Results

We examined how prosocial and financial incentives influenced the types of species posted. The app categorized posted species into groups, including rare species (e.g., endangered and quasi-endangered species and species with limited information), non-rare species, and invasive species. The results revealed that the prosocial incentive increased posts on rare species. In contrast, the financial incentive tended to boost posts of less rare and invasive species.

---

[11] Table Appendix B.1 addresses the concern that participants may have been taking and saving photos of species on weekends and posting them during the weekdays to maximize their financial rewards. If this was the case, the number of weekend postings should has decreased in the financial incentive treatment group; however, we did not observe this phenomenon.

**Table 4.** Further Analysis: Effects on Contents of Posts

|  | (1) | (2) | (3) | (4) | (5) | (6) |
|---|---|---|---|---|---|---|
| Analysis sample: | Total | | | | | |
| Number of participants: | n=830 | | | | | |
| Number of observations: | n=2,490 | | | | | |
| Dependent variable: | Number of posts on rare species | | Number of posts on less rare species | | Number of posts on invasive species | |
| Prosocial incentive*Treatment week 1 | 2.084* | 2.068* | 0.860 | 0.863 | 0.020 | 0.019 |
|  | (1.246) | (1.239) | (0.928) | (0.925) | (0.120) | (0.119) |
| Prosocial incentive*Treatment week 2 | 0.824** | 0.823** | -0.395 | -0.376 | -0.246 | -0.247 |
|  | (0.371) | (0.368) | (0.906) | (0.912) | (0.250) | (0.251) |
| Financial incentive*Treatment week 1 | 0.654 | 0.602 | 1.496* | 1.473* | 0.250** | 0.246** |
|  | (0.444) | (0.446) | (0.778) | (0.778) | (0.116) | (0.116) |
| Financial incentive*Treatment week 2 | 0.215 | 0.141 | 0.699 | 0.680 | -0.020 | -0.029 |
|  | (0.418) | (0.425) | (0.911) | (0.920) | (0.246) | (0.249) |
| Treatment week 1 | -0.775** | -5.263 | -0.611 | -4.099 | -0.013 | -0.347 |
|  | (0.304) | (5.108) | (0.500) | (3.348) | (0.068) | (0.439) |
| Treatment week 2 | -0.789*** | -6.625 | -0.375 | -4.968 | 0.147 | -0.275 |
|  | (0.281) | (6.645) | (0.802) | (4.432) | (0.240) | (0.622) |
| Time-varing attributes | NO | YES | NO | YES | NO | YES |
| Constant term | 1.304*** | 15.073 | 4.320*** | 14.554 | 0.356*** | 1.316 |
|  | (0.167) | (14.583) | (0.188) | (9.295) | (0.034) | (1.219) |

Notes: We use standard errors clustered at each participant; * $p < 0.1$, ** $p < 0.05$, *** $p < 0.01$

**Table 4, Column 1** presents the results for rare species posts. The prosocial incentive group showed a coefficient of 2.084 in the first treatment week, which was weakly statistically significant, at the 10% level; the coefficient in the second treatment week was 0.824, which was smaller than for week 1 but achieved statistical significance at the 5% level. The coefficients for the financial incentive were small for both weeks and not statistically significant.

**Columns 3 and 5** present the results for less rare and invasive species posts. The financial incentive increased both indicators only during the first treatment week. The coefficient for the number of less rare species posts was 1.496 and statistically significant, albeit at the 10% level. The coefficient for the number of invasive species posts was 0.250 and statistically significant at the 5% level.

For robustness check, **columns 2, 4, and 6** additionally controlled for weather data (weekly average temperature, rainfall, and sunshine hours) for the participants' locations. The results showed that the size and statistical significance remained largely unchanged.

### 4.2. Interpretation

The result of **Table 4, Column 1** suggests a substantial impact of the prosocial incentive on the number of posted photos of rare species. Initially, its number averaged 1.304, according to the constant term. During the first treatment week, there was a decrease in the control group from 1.304 to 0.515 (a drop of 0.789). The prosocial incentive reversed this trend, increasing the number of postings by around 0.824 (from 0.515 to 1.339), or a substantial rise of 160.0%.

**Tables 2 and 4** show that although the number of posted photos of unique, different species in the prosocial incentive treatment group did not increase, the proportion of rare species photos increased. This result indicates the possibility that the increase in rare species posts may not be accidental due to taking and posting more photos. The participants likely knew the locations of rare species, and the prosocial incentive encouraged them to visit these locations and photograph the species.

The financial incentive treatment effect on the number of posted photos of less rare and invasive species was substantial. For example, the effect size for invasive species posts means to add 0.250 to the average number of posts of 0.356 at the baseline period, with an increase of +70.2%.

**Sections 4 and 5** suggests that the participants in the financial incentive treatment group likely posted more photos of less rare species that are often encountered. This indicates a behavioral pattern of taking and posting photos of species in their vicinity to maximize monetary rewards. However, it should be noted that this financial incentive increased the number of posted photos of invasive species, which could substantially influence eco-systems. As the participants posted more photos of familiar, less rare species, they may have unintentionally encountered and submitted photos of invasive species. As the baseline mean of posted photos of these species was relatively low at 0.356, this increase could be meaningful.

## 5. Discussion, Limitations, and Conclusions

This study proposed two interventions to encourage people to report plant, insect, or animal species information, focusing on the conditions where posts typically decrease. One unique feature of our study was employing prosocial incentives, in which donations were made to biodiversity conservation activities in response to people' posts, in addition to financial incentives. We evaluated the effectiveness of these two interventions through a randomized controlled trial with 830 users of a Japanese smartphone app widely-used for posting species photos. The prosocial incentive did not increase the average number of species posts, whereas the financial incentive significantly increased the number of these posts. However, we found another tendency when looking at contents of the posted species information. The prosocial incentive increased the proportion of posted photos of rare species. In contrast, the financial incentive increased the number of photos of less rare and invasive species.

Our findings indicate that the prosocial and financial incentives could stimulate different motivations and encourage different posting behaviors. This suggests developing policies using different incentives schemes depending on the species for which governments and practitioners need to collect information. This also suggests a concern that employing a single incentive could lead to the collection of unwanted information. This concern has been raised by Diekert et al. (2023), who found that financial incentives increased the number of species information posts but decreased the diversity of the posted species information. Although our results on financial incentives were similar to theirs, our findings highlight the potential of financial incentives to increase the number of posts on invasive species that could significantly impact the ecosystem, in addition to less rare species postings. Balancing the implementation of both prosocial and financial incentives could lead to more comprehensive biodiversity information collection.

In addition, our study contributes to advocating for behavioral intervention studies in biodiversity conservation. In this field, there have been limited cases of field experiments on conservation of threatened species (Walsh, 2021), water use management (Ferraro, 2011), and fundraising activities (Kubo et al., 2018; Kubo et al., 2023), partly due to the complexity of biodiversity metrics (Baylis et al, 2016). Exploring causal influence of financial and non-financial interventions through field experiments beyond the limited topics has significant potential to provide policy and practical implications for biodiversity conservation. Further studies should be encouraged to integrate ecological knowledge and findings of behavioral change research in other important topics, including food choice, transportation, recreation, wildlife trade, education, land use change, etc. (The Behavioral Insights Team, 2019).

This is the first study to conduct a field experiment exploring the impacts of prosocial and financial interventions to enhance citizen science in biodiversity conservation. However, this study has several limitations. This study does not elucidate to what extent the treatments could reduce the bias of citizen-posted species information, what mechanisms generate each treatment effect, and to what extent the effect changes when users repeatedly receive the same incentive. Furthermore, the concern of external validity remains, because we obtained participants' opt-in informed consent and used their data for analysis. These questions should be examined in future studies. Nevertheless, intervention studies on voluntary information provision by citizens for biodiversity conservation are an emerging line of research. Our study makes significant academic and policy contributions by providing new insights to this emerging literature using a controlled field experiment.

**Acknowledgements**

First of all, we thank Shojiro Fujiki, CEO of Biome, Inc. and its staff for their warm support and cooperation in using their smartphone app as a field for our randomized controlled trial. We also thank the participants at Japan Conference for 2030 Global Biodiversity Framework by the Ministry of the Environment, the 26th Experimental Social Science Conference, and the 17th Annual Meeting of the Association for Behavioral Economics and Finance for their helpful comments.

**Declaration of Generative AI and AI-assisted technologies in the writing process**

During the preparation of this work the authors used DeepL and ChatGPT to improve the readability and proofreading of the English text we have written. After using this tool, the authors reviewed and edited the content as needed and take full responsibility for the content of the manuscript.

# References

Adena, M., Huck, S., 2017. Matching donations without crowding out? Some theoretical considerations, a field, and a lab experiment. J. Public Econ. 148, 32–42.

Allcott, H., 2011. Social norms and energy conservation. J. Public Econ. 95, 1082–1095.

Atsumi, K., Nishida, Y., Ushio, M., Nishi, H., Genroku, T., Fujiki, S., 2023. Boosting biodiversity monitoring using smartphone-driven, rapidly accumulating community-sourced data. bioRxiv 2023.09.13.557657.

Barber, A., West, J., 2022. Conditional cash lotteries increase COVID-19 vaccination rates. J. Health Econ. 81, 102578.

Barte, J.C.M., Wendel-Vos, G.C.W., 2017. A systematic review of financial incentives for physical activity: the effects on physical activity and related outcomes. Behav. Med. 43, 79–90.

Bas, Y., Devictor, V., Moussus, J.P., Jiguet, F., 2008. Accounting for weather and time-of-day parameters when analysing count data from monitoring programs. Biodivers. Conserv. 17, 3403–3416.

Baylis, K., Honey-Rosés, J., Börner, J., Corbera, E., Ezzine-de-Blas, D., Ferraro, P.J., Lapeyre, R., Persson, U.M., Pfaff, A. and Wunder, S., 2016. Mainstreaming impact evaluation in nature conservation. Conserv. Lett. 9(1), pp.58–64.

The Behavioral Insights Team, 2019. Behavior Change for Nature: A Behavioral Science Toolkit for Practitioners. https://www.bi.team/wp-content/uploads/2019/04/2019-BIT-Rare-Behavior-Change-for-Nature-digital.pdf (accessed 25 February 2024).

Bénabou, R., Tirole, J., 2003. Intrinsic and extrinsic motivation. Rev. Econ. Studies. 70, 489–

520.

Bird, T.J., Bates, A.E., Lefcheck, J.S., Hill, N.A., Thomson, R.J., Edgar, G.J., Stuart-Smith, R.D., Wotherspoon, S., Krkosek, M., Stuart-Smith, J.F., Pecl, G.T., Barrett, N., Frusher, S., 2014. Statistical solutions for error and bias in global citizen science datasets. Biol. Conserv. 173, 144–154.

Bradter, U., Mair, L., Jönsson, M., Knape, J., Singer, A., Snäll, T., 2018. Can opportunistically collected Citizen Science data fill a data gap for habitat suitability models of less common species? Methods Ecol. Evol. 9, 1667–1678.

Brum-Bastos, V.S., Long, J.A., Demšar, U., 2018. Weather effects on human mobility: a study using multi-channel sequence analysis. Comput. Environ. Urban Syst. 71, 131–152.

Cameron, T.A., Kolstoe, S.H., 2022. Using auxiliary population samples for sample-selection correction in models based on crowd-sourced volunteered geographic information. Land Econ. 98, 1–21.

Cornes, R., Sandler, T., 1996. The Theory of Externalities, Public Goods, and Club Goods. Cambridge University Press, Cambridge.

Courter, J.R., Johnson, R.J., Stuyck, C.M., Lang, B.A., Kaiser, E.W., 2013. Weekend bias in Citizen Science data reporting: implications for phenology studies. Int. J. Biometeorol. 57, 715–720.

DellaVigna, S., Linos, E., 2022. RCTs to scale: comprehensive evidence from two nudge units. Econometrica. 90, 81–116.

Diekert, F., Munzinger, S., Schulemann-Maier, G., Städtler, L., 2023. Explicit incentives increase citizen science recordings. Conserv. Lett. 16, e12973.

Eichenberg, D., Bowler, D.E., Bonn, A., Bruelheide, H., Grescho, V., Harter, D., Jandt, U., May, R., Winter, M., Jansen, F., 2020. Widespread decline in Central European plant diversity across six decades. Glob. Chang. Biol. 27, 1097–1110.

Epperson, R., Reif, C., 2019. Matching subsidies and voluntary contributions: a review. J. Econ. Surv. 33, 1578–1601.

Ferraro, P.J., Miranda, J.J., Price, M.K. 2011. The persistence of treatment effects with norm-based policy instruments: evidence from a randomized environmental policy experiment. Amer. Econ. Rev. 101(3), 318–322.

Galárraga, O., Bohlen, L.C., Dunsiger, S.I., Lee, H.H., Emerson, J.A., Boyle, H.K., Strohacker, K., Williams, D.M., 2020. Small sustainable monetary donation-based incentives to promote physical activity: A randomized controlled trial. Health Psychol. 39, 265–268.

Gneezy, U., Meier, S., Rey-Biel, P., 2011. When and why incentives (don't) work to modify behavior. J. Econ. Perspect. 25, 191–210.

Handberg, Ø.N., Angelsen, A., 2019. Pay little, get little; pay more, get a little more: a framed forest experiment in Tanzania. Ecol. Econ. 156, 454–467.

Harkins, K.A., Kullgren, J.T., Bellamy, S.L., Karlawish, J., Glanz, K., 2017. A trial of financial and social incentives to increase older adults' walking. Am. J. Prev. Med. 52, e123–e130.

Harrison, G.W., List, J.A., 2004. Field experiments. J. Econ. Lit. 42, 1009–1055.

Imas, A., 2014. Working for the "warm glow": on the benefits and limits of prosocial incentives. J. Public Econ. 114, 14–18.

IPBES, 2024. The Intergovernmental Science-Policy Platform on Biodiversity and Ecosystem Services. https://www.ipbes.net/ (accessed 31 January 2024).

Kerr, J., Vardhan, M., Jindal, R., 2012. Prosocial behavior and incentives: evidence from field experiments in rural Mexico and Tanzania. Ecol. Econ. 73, 220–227.

Kubo, T., Shoji, Y., Tsuge, T., Kuriyama, K., 2018. Voluntary contributions to hiking trail maintenance: evidence from a field experiment in a National Park, Japan. Ecol. Econ. 144, 124–128.

Kubo, T., Yokoo, H.F., Veríssimo, D., 2023. Conservation fundraising: evidence from social media and traditional mail field experiments. Conserv. Lett. 16, e12931.

Liu, G., Kingsford, R.T., Callaghan, C.T., Rowley, J.J.L., 2022. Anthropogenic habitat modification alters calling phenology of frogs. Glob. Chang. Biol. 28, 6194–6208.

Negrete, L., Lenguas Francavilla, M., Damborenea, C., Brusa, F., 2020. Trying to take over the world: potential distribution of Obama nungara (Platyhelminthes: Geoplanidae), the Neotropical land planarian that has reached Europe. Glob. Chang. Biol. 26, 4907–4918.

Perrings, C., Gadgil, M., 2003. Conserving biodiversity: reconciling local and global public benefits, in: Kaul, I. (Ed.), Providing Global Public Goods: Managing Globalization. Oxford University Press, New York, pp. 532–555.

Samek, A., 2019. Advantages and disadvantages of field experiments, in: Schram, A., Ule, A. (Eds.), Handbook of Research Methods and Applications in Experimental Economics. Edward Elgar Publishing Limited, Cheltenhampp, pp. 104–120.

Sasaki, S., Kubo, T., 2022. Prosocial Incentive versus Financial Incentive in Biodiversity Conservation: A Field Experiment. AEA RCT Registry.

https://doi.org/10.1257/rct.10210-2.1.

Sasaki, S., Kurokawa, H., Ohtake, F., 2022. An experimental comparison of rebate and matching in charitable giving: the case of Japan. Jpn. Econ. Rev. 73, 147–177.

Scharf, K., Smith, S., Ottoni-Wilhelm, M., 2022. Lift and shift: the effect of fundraising interventions in charity space and time. Am. Econ. J. Econ. Policy. 14, 296–321.

Sparks, T.H., Huber, K., Tryjanowski, P., 2008. Something for the weekend? Examining the bias in avian phenological recording. Int. J. Biometeorol. 52, 505–510.

Sumida, Y., Yoshikawa, T., Tanaka, S., Taketani, H., Kanemasa, K., Nishimura, T., Yamaguchi, K., Mitsuyoshi, H., Yasui, K., Minami, M., Naito, Y., Itoh, Y., 2014. The 'donations for decreased ALT (D4D)' prosocial behavior incentive scheme for NAFLD patients. J. Public Health 36, 629–634.

Van Strien, A.J., Van Swaay, C.A.M., Termaat, T., 2013. Opportunistic citizen science data of animal species produce reliable estimates of distribution trends if analysed with occupancy models. J. Appl. Ecol. 50, 1450–1458.

Walsh, P.J. 2021. Behavioural approaches and conservation messages with New Zealand's threatened kiwi. Glob. Ecol. Conserv. 28, e01694.

Yuan, Y., Nicolaides, C., Pentland, A., Eckles, D., 2021. Promoting physical activity through prosocial incentives on mobile platforms. http://dx.doi.org/10.2139/ssrn.3960420 (accessed 31 January 2024).

**Appendix A. Experimental Schedule**

Recruitment with Baseline Survey: September 15–30
↓
Information Provision by Groups (1): October 11
↓
<u>First Treatment Week: October 17–21</u>
Information Provision by Groups (2): October 17
↓
<u>Second Treatment Week: October 24–28</u>
Information Provision by Groups (3): October 24
↓
First Post-Treatment Week: October 31–November 04
↓
Second Post-Treatment Week: November 07–11

*Note*: To account for fairness among participants, we offered those in the control group the opportunity to receive a prosocial or financial incentive after November 8. Therefore, we did not use data after this date in the analysis.

# Appendix B

**Table Appendix B.1.** Effects on Number of Posts on Weekends During the Treatment Period

|  | (1) | (2) | (3) | (4) |
|---|---|---|---|---|
| Analysis sample: | \multicolumn{4}{c}{Total} | | | |
| Number of participants: | \multicolumn{4}{c}{n=830} | | | |
| Number of observations: | \multicolumn{4}{c}{n=2,490} | | | |
| Dependent variable: | Number of posts of unique, different species photos on weekends | | Number of posts of species photos on weekends | |
| Prosocial incentive*Treatment week 1 | -0.656 | -0.660 | -0.617 | -0.615 |
|  | (0.604) | (0.604) | (0.695) | (0.696) |
| Prosocial incentive*Treatment week 2 | -0.435 | -0.445 | -0.370 | -0.374 |
|  | (0.572) | (0.571) | (0.630) | (0.630) |
| Financial incentive*Treatment week 1 | 0.393 | 0.402 | 0.474 | 0.488 |
|  | (0.684) | (0.685) | (0.788) | (0.789) |
| Financial incentive*Treatment week 2 | 0.256 | 0.274 | 0.208 | 0.233 |
|  | (0.615) | (0.612) | (0.679) | (0.676) |
| Treatment week 1 | -0.609 | 0.937 | -0.676 | 0.743 |
|  | (0.473) | (1.693) | (0.562) | (1.984) |
| Treatment week 2 | -1.402*** | 0.595 | -1.520*** | 0.300 |
|  | (0.481) | (2.176) | (0.525) | (2.510) |
| Time-varing attributes | NO | YES | NO | YES |
| Constant term | 3.578*** | -0.608 | 4.122*** | 0.138 |
|  | (0.147) | (4.734) | (0.163) | (5.424) |

Notes: We use standard errors clustered at each participant; * p < 0.1, ** p < 0.05, *** p < 0.01

**Table Appendix B.2.** Effects on Number of Posts After the Treatment Period

|  | (1) | (2) | (3) | (4) |
|---|---|---|---|---|
| Analysis sample: | \multicolumn{4}{c}{Total} | | | |
| Number of participants: | \multicolumn{4}{c}{n=830} | | | |
| Number of observations: | \multicolumn{4}{c}{n=2,490} | | | |
| Dependent variable: | Number of posts of unique, different species photos | | Number of posts of species photos | |
| Prosocial incentive*Post-Treatment week 1 | 0.617 | 0.598 | 0.457 | 0.427 |
|  | (0.863) | (0.861) | (0.994) | (0.992) |
| Prosocial incentive*Post-Treatment week 2 | 1.048 | 1.007 | 1.259 | 1.201 |
|  | (0.840) | (0.836) | (0.943) | (0.936) |
| Financial incentive*Post-Treatment week 1 | 0.739 | 0.730 | 0.207 | 0.190 |
|  | (0.930) | (0.928) | (1.057) | (1.055) |
| Financial incentive*Post-Treatment week 2 | 0.773 | 0.770 | 0.530 | 0.524 |
|  | (0.883) | (0.881) | (0.964) | (0.961) |
| Post-Treatment week 1 | -0.765 | 3.351 | -0.400 | 4.716** |
|  | (0.720) | (2.081) | (0.834) | (2.312) |
| Post-Treatment week 2 | -1.980*** | 2.655 | -1.844** | 3.953 |
|  | (0.675) | (2.256) | (0.741) | (2.558) |
| Time-varing attributes | NO | YES | NO | YES |
| Constant term | 5.304*** | -3.730 | 6.046*** | -4.874 |
|  | (0.260) | (5.214) | (0.316) | (5.819) |

Notes: We use standard errors clustered at each participant; * $p < 0.1$, ** $p < 0.05$, *** $p < 0.01$

# Supplementary Materials on Information Provision by Groups

**C1: E-mail to Control Group on 11 October 2022**

Subject: [biome] Important Notice for the Posting Marathon Campaign

Thank you for participating in the Posting Marathon Campaign!
The campaign office is pleased to send you the first e-mail.

The campaign period is from Tuesday, October 11 to Wednesday, November 30.
Take and post photos of plant, insect, or animal species in your spare time on weekdays during this period!

You can take photos
- on your way to work or school
- during lunch or break time, or
- while walking

For example, if you look at the foot of street trees, you may find rare species!

The campaign office will e-mail you various information once a week.
The next e-mail will be on Monday, October 17. Don't miss it!

----
The campaign office

**C2: E-mail to Control Group on 17 October 2022**

Subject: [biome] The Second E-mail for the Posting Marathon Campaign

Thank you for participating in the Posting Marathon Campaign!
The campaign office is pleased to send you the second e-mail. How many photos of plant, insect, or animal species did you take and post last week?

The campaign period is from Tuesday, October 11 to Wednesday, November 30.
Take and post photos of plant, insect, or animal species in your spare time on weekdays during this period!

[Biodiversity News: Do you know what "biodiversity" means?]
It is estimated that Earth millions of species. Biodiversity is a phenomenon in which species with various characteristics live while supporting each other.

According to the Japanese government's "Public Opinion Survey on Environmental Issues (August 2019)," ## % of respondents answered that they knew the meaning of "biodiversity"!
*You can find the answer here => https://survey.gov-online.go.jp/r01/r01-kankyou/2-2.html

The campaign office will e-mail you various information once a week.
The next e-mail will be on Monday, October 24. Don't miss it!

----
The campaign office

**C3: E-mail to Control Group on 24 October 2022**

Subject: [biome] The Third E-mail for the Posting Marathon Campaign

Thank you for participating in the Posting Marathon Campaign!
The campaign office is pleased to send you the third e-mail. How many photos of plant, insect, or animal species did you take and post last week?

The campaign period is from Tuesday, October 11 to Wednesday, November 30.
Take and post photos of plant, insect, or animal species in your spare time on weekdays during this period!

[Biodiversity News: The latest survey results are available now!]
According to the Japanese government's "Public Opinion Survey on Environmental Issues (August 2019)," 20.1% of respondents answered that they knew the meaning of "biodiversity"!

The latest survey was released on October 14, 2022! The percentage of respondents who answered that they knew the meaning of "biodiversity" has changed dramatically. Do you think it has risen or fallen?
*You can find the answer here => https://survey.gov-online.go.jp/hutai/r04/r04-seibutsutayousei/index.html

----
The campaign office

**P1: E-mail to Prosocial Incentive Treatment Group on 11 October 2022**

Subject: [biome] Important Notice for the Posting Marathon Campaign

Thank you for participating in the Posting Marathon Campaign!
The campaign office is pleased to send you the first e-mail.

The campaign period is from Tuesday, October 11 to Wednesday, November 30.
Take and post photos of plant, insect, or animal species in your spare time on weekdays during this period!

You can take photos
- on your way to work or school
- during lunch or break time, or
- while walking

For example, if you look at the foot of street trees, you may find rare species!

[Special Event]
Take and post "photos of 10 unique, different species" every weekday next week, October 17–21!

For each different species photo you post during these five weekdays, the campaign office will donate 25 JPY to the endangered species protection activities. If you post photos of 10 different species, the office will donate a maximum of 250 JPY.

The donation will be delivered to the Japan Wildlife Conservation Society's "Activities to Protect Endangered Species in Japan." => https://www.nacsj.or.jp/

Join the activities to protect endangered species by posting species!
A similar event will be held from October 24 to 28.

The campaign office will e-mail you various information once a week.
The next e-mail will be on Monday, October 17. Don't miss it!

----
The campaign office

**P2: E-mail to Prosocial Incentive Treatment Group on 17 October 2022**

Subject: [biome] Posting Marathon Campaign's "Special Event"

Thank you for participating in the Posting Marathon Campaign!
The campaign office is pleased to send you the second e-mail. How many photos of plant, insect, or animal species did you take and post last week?

The campaign period is from Tuesday, October 11 to Wednesday, November 30.
Take and post photos of plant, insect, or animal species in your spare time on weekdays during this period!

["Special Event" starts today!]
Take and post "photos of 10 unique, different species" every weekday this week, October 17–21!

For each different species photo you post during these five weekdays, the campaign office will donate 25 JPY to the endangered species protection activities. If you post photos of 10 different species, the office will donate a maximum of 250 JPY.

The donation will be delivered to the Japan Wildlife Conservation Society's "Activities to Protect Endangered Species in Japan." => https://www.nacsj.or.jp/

Join the activities to protect endangered species by posting their photos!
A similar event will be held from October 24 to 28.

[Biodiversity News: Do you know what "biodiversity" means?]
It is estimated that Earth millions of species. Biodiversity is a phenomenon in which species with various characteristics live while supporting each other.

According to the Japanese government's "Public Opinion Survey on Environmental Issues (August 2019)," ## % of respondents answered that they knew the meaning of "biodiversity"!
*You can find the answer here => https://survey.gov-online.go.jp/r01/r01-kankyou/2-2.html

The campaign office will e-mail you various information once a week.

The next e-mail will be on Monday, October 24. Don't miss it!
----
The campaign office

**P3: E-mail to Prosocial Incentive Treatment Group on 24 October 2022**

Subject: [biome] Posting Marathon Campaign's "Special Event" (5 days left!)

Thank you for participating in the Posting Marathon Campaign!
The campaign office is pleased to send you the second e-mail. How many photos of plant, insect, or animal species did you take and post last week?

The campaign period is from Tuesday, October 11 to Wednesday, November 30.
Take and post photos of plant, insect, or animal species in your spare time on weekdays during this period!

["Special Event" continues this week!]
Take and post "photos of 10 unique, different species" every weekday this week, October 24–28!

For each different species photo you post during these five weekdays, the campaign office will donate 25 JPY to the endangered species protection activities. If you post photos of 10 different species, the office will donate a maximum of 250 JPY.

The donation will be delivered to the Japan Wildlife Conservation Society's "Activities to Protect Endangered Species in Japan." => https://www.nacsj.or.jp/

Join the activities to protect endangered species by posting their photos!
This event will end this week.

[Biodiversity News: The latest survey results are available now!]
According to the Japanese government's "Public Opinion Survey on Environmental Issues (August 2019)," 20.1% of respondents answered that they knew the meaning of "biodiversity"!

The latest survey was released on October 14, 2022! The percentage of respondents who answered that they knew the meaning of "biodiversity" has changed dramatically. Do you think it has risen or fallen?
*You can find the answer here => https://survey.gov-online.go.jp/hutai/r04/r04-seibutsutayousei/index.html

----
The campaign office

**F1: E-mail to Financial Incentive Treatment Group on 11 October 2022**

Subject: [biome] Important Notice for the Posting Marathon Campaign

Thank you for participating in the Posting Marathon Campaign!
The campaign office is pleased to send you the first e-mail.

The campaign period is from Tuesday, October 11 to Wednesday, November 30.
Take and post photos of plant, insect, or animal species in your spare time on weekdays during this period!

You can take photos
- on your way to work or school
- during lunch or break time, or
- while walking

For example, if you look at the foot of street trees, you may find rare species!

[Special Event]
Take and post "photos of 10 unique, different species" every weekday next week, October 17–21!

For each different species photo you post during these five weekdays, you will receive an Amazon gift certificates (e-mail type) of 25 JPY. If you post photos of 10 different species, you will receive a maximum of 250 JPY gift certificates.

Amazon gift certificates can be used for online shopping at Amazon. =>
https://www.amazon.co.jp/

Post species photos and enjoy shopping!
A similar event will be held from October 24 to 28.

The campaign office will e-mail you various information once a week.
The next e-mail will be on Monday, October 17. Don't miss it!

----
The campaign office

**F2: E-mail to Financial Incentive Treatment Group on 17 October 2022**

Subject: [biome] Posting Marathon Campaign's "Special Event"

Thank you for participating in the Posting Marathon Campaign!
The campaign office is pleased to send you the second e-mail. How many photos of plant, insect, or animal species did you take and post last week?

The campaign period is from Tuesday, October 11 through Wednesday, November 30. Take and post photos of plant, insect, or animal species in your spare time on weekdays during this period!

["Special Event" starts today!]
Take and post "photos of 10 unique, different species" every weekday this week, October 17–21!

For each different species photo you post during these five weekdays, you will receive an Amazon gift certificates (e-mail type) of 25 JPY. If you post photos of 10 different species, you will receive a maximum of 250 JPY gift certificates.

Amazon gift certificates can be used for online shopping at Amazon. => https://www.amazon.co.jp/

Post species photos and enjoy shopping!
A similar event will be held from October 24 to 28.

[Biodiversity News: Do you know what "biodiversity" means?]
It is estimated that Earth millions of species. Biodiversity is a phenomenon in which species with various characteristics live while supporting each other.

According to the Japanese government's "Public Opinion Survey on Environmental Issues (August 2019)," ## % of respondents answered that they knew the meaning of "biodiversity"!
*You can find the answer here => https://survey.gov-online.go.jp/r01/r01-kankyou/2-2.html

The campaign office will e-mail you various information once a week.

The next e-mail will be on Monday, October 24. Don't miss it!
----
The campaign office

**P3: E-mail to Financial Incentive Treatment Group on 24 October 2022**

Subject: [biome] Posting Marathon Campaign's "Special Event" (5 days left!)

Thank you for participating in the Posting Marathon Campaign!
The campaign office is pleased to send you the third e-mail. How many photos of plant, insect, or animal species did you take and post last week?

The campaign period is from Tuesday, October 11 through Wednesday, November 30. Take and post photos of plant, insect, or animal species in your spare time on weekdays during this period!

["Special Event" continues this week!]
Take and post "photos of 10 unique, different species" every weekday this week, October 24–28!

For each different species photo you post during these five weekdays, you will receive an Amazon gift certificates (e-mail type) of 25 JPY. If you post photos of 10 different species, you will receive a maximum of 250 JPY gift certificates.

Amazon gift certificates can be used for online shopping at Amazon. =>
https://www.amazon.co.jp/

Post species photos and enjoy shopping!
This event will end this week.

[Biodiversity News: The latest survey results are available now!]
According to the Japanese government's "Public Opinion Survey on Environmental Issues (August 2019)," 20.1% of respondents answered that they knew the meaning of "biodiversity"!

The latest survey was released on October 14, 2022! The percentage of respondents who answered that they knew the meaning of "biodiversity" has changed dramatically. Do you think it has risen or fallen?
*You can find the answer here => https://survey.gov-online.go.jp/hutai/r04/r04-seibutsutayousei/index.html

----
The campaign office